\documentclass[twocolumn,showpacs,preprintnumbers]{revtex4}
\usepackage{amsmath}
\usepackage{graphicx}
\usepackage{dcolumn}
\usepackage{bm}

\begin{document}

\title{ Improved Silbey-Harris polaron ansatz for the   spin-boson model}

\author{Shu He$^{1,2}$, Liwei Duan$^{1}$, and Qing-Hu Chen$^{1,3,*}$}
\address{
$^{1}$ Department of Physics, Zhejiang University, Hangzhou 310027,
 China \\
$^{2}$  Department of Physics and Electronic Engineering,  Sichuan Normal University, Chengdu 610066, China\\
$^{3}$  Collaborative Innovation Center of Advanced Microstructures, Nanjing 210093, China
 }
\date{\today }

\begin{abstract}
In this paper, the well-known Silbey-Harris (SH) polaron  ansatz for the spin-boson model  is improved by adding  orthogonal displaced Fock states. The  obtained results for the ground-state in  all baths converge very quickly within finite displaced Fock states and corresponding SH results are corrected considerably.   Especially for the sub-Ohmic spin-boson model, the  converging results are obtained for $0<s<1/2$ in the fourth-order correction and very accurate  critical coupling strengths of the quantum phase transition are achieved. Converging magnetization in the biased spin-boson model is also arrived at. Since the present improved SH ansatz can yield very accurate, even almost exact results, it should have wide applications and extensions  in various spin-boson model and related fields.

\end{abstract}

\pacs{03.65.Yz, 03.65.Ud, 71.27.+a, 71.38.k}
\maketitle

\section{Introduction}

The celebrated spin-boson model ~\cite{Leggett,weiss} describes a qubit
(two-level system) coupled with a dissipative environment represented by a
bath of continuous bosonic modes. There is currently considerable interest
in this quantum many-body system due to the rich physics of quantum
criticality and decoherence~\cite{weiss,Hur,Kopp}, applications to the emerging
field of quantum computations~\cite{Thorwart}, quantum devices~\cite{mak},
and quantum biology~\cite{reng,omer}. It is widely used to study the
microscopic behavior of  open quantum systems~\cite{Leggett}. The
coupling between the qubit and the environment is characterized by a
spectral function $J(\omega )$ which is proportional to $\omega ^s$. The
spectral exponent $s$ varies the coupling into three different cases:
sub-Ohmic ($s<1$), Ohmic ($s=1$), and super-Ohmic ($s>1$).

As a paradigmatic model to study the influence of environment on the quantum
system, the spin-boson model has been extensively and persistently studied
by many analytical and numerical approaches. On the analytical side, a
pioneer work is undoubtedly the variational study based on the polaronic
unitary transformation also known as Silbey-Harris ansatz (SH) ansatz ~\cite{Silbey}. A similar
analytical approach was developed to study both static and dynamical
behavior of the dissipative two-level system \cite{zheng}. The original
symmetric SH ansatz using a single coherent state was generalized to the
asymmetric form for the sub-Ohmic baths~\cite{Chin}. Recently, the
asymmetric SH ansatz was modified by superposing more than one nonorthogonal
coherent states on the equal footing~\cite{mD1,Blunden,ZhengLu}, and the
equilibrium reduced density matrix in the SH frames was corrected to the
second-order ~\cite{cao,Duan}.

On the numerical side, almost all advanced numerical approaches in the
quantum many-particle physics have been applied and extended to this model
and many interesting results have been obtained. The numerical
renormalization group  was applied at the earlier stage~\cite{Bulla} for
the sub-Ohmic baths, yielding non-mean-field  critical
exponents of the quantum phase transitions (QPT) for $0<s<1/2$ due to the
Hilbert-space truncation error and the mass flow error~\cite{vojta,phi,Tong}.
Later on, quantum Monte Carlo (QMC) simulations based on an imaginary time path
integral~\cite{QMC}, sparse polynomial space approach~\cite{ED}, and exact
diagonalization in terms of shift bosons \cite{Zhang} have sequentially
developed and all found the mean-field critical exponent for $0<s<1/2$. The
density matrix renormalization group  was also applied, but not
successful in the analysis of the critical phenomena~\cite{DMRG}. More
recently, using the density matrix renormalization group algorithm combined with the optimized phonon basis,
a variational matrix product state  approach formulated on a Wilson
chain~\cite{VMPS} was developed and the Hilbert-space truncation can be
alleviated systematically. An alternative to the conventional
matrix product state representation was also proposed~\cite{Frenzel}. Most recently, a highly
efficient  numerical method based on a time-dependent variational principle
for matrix product states \cite{Wall,Schreoer,Haegeman} was proposed and extended to simulate   quantum dynamics of the spin-boson model.

Analytical exact study should be very challenging in the spin-boson model
due to the infinite modes of the baths, unlike the single mode case~\cite%
{Chen2011,Braak, Chen2012}. Even the well controlled and fast converging
analytical study was still lacking until now. In this paper, we propose an
analytical approach for the ground state (GS) of the spin-boson model anchored
to SH polaron ansatz. The bosonic state is expanded in the orthogonal
displaced Fock states basis so that the SH results can be improved systematically. The  high order displaced Fock states can essentially include the
many-body correlations for bosons, which are obviously lacking in the SH wavefunction, so  the results with increasing orders should approach the exact ones asymptotically.

The paper is organized as follows. In Sec. II, we briefly introduce the spin-boson
model,  and describe our improved SH ansatz in terms of displaced Fock states in detail. In addition, we  test the approach in the single mode case.
In Sec. III, we apply it to the spin-boson model for different baths,  and  further extend it to the
biased spin-boson model. A short summary is given in Sec.IV.

\section{Model and Displaced Fock states}

The Hamiltonian of the spin-boson model is given by
\begin{equation}
H=-\frac \Delta 2\sigma _x+\sum_k\omega _ka_k^{\dagger }a_k+\frac 12\sigma
_z\sum_kg_k(a_k^{\dagger }+a_k),  \label{hamiltonian}
\end{equation}
where $\sigma _x$ and $\sigma _z$ are Pauli matrices, $\Delta $ is the
tunneling amplitude between two levels, $\omega _k$ and $a_k^{\dagger }$ are
the frequency and creation operator of the $k$-th harmonic oscillator, and $%
g_k$ is the interaction strength between the $k$-th bosonic mode and the
local spin. The spin-boson coupling is characterized by the spectral
function,
\begin{equation}
J(\omega )=\pi \sum_kg_k^2\delta (\omega _k-\omega )=2\pi \lambda \omega
_c^{1-s}\omega ^s,0<\omega <\omega _c,
\end{equation}
with $\omega _c$ a cutoff frequency. The dimensionless parameter $\lambda $
denotes the coupling strength.

To outline the approach more intuitively, we first consider the case without
symmetry breaking. By using $|\uparrow \rangle $ and $|\downarrow \rangle $
to represent the eigenstate of $\sigma _z,$ the GS wavefunction can be in
principle expressed in the following standard set of complete orthogonal basis $%
\prod_{i=0}^na_{k_i}^{\dagger }|0\rangle $ in Fock space

\begin{eqnarray}
&&|\Psi ^{\prime }\rangle =\left( 1+\sum_{k}\alpha _{k}a_{k}^{\dagger
}+\sum_{k_{1},k_{2}}u_{k_{1},k_{2}}a_{k_{1}}^{\dagger }a_{k_{2}}^{\dagger
}+...\right) |0\rangle |\uparrow \rangle  \notag \\
&&+\left( 1-\sum_{k}\alpha _{k}a_{k}^{\dagger
}+\sum_{k_{1},k_{2}}u_{k_{1},k_{2}}a_{k_{1}}^{\dagger }a_{k_{2}}^{\dagger
}+...\right) |0\rangle |\downarrow \rangle ,  \label{Fock}
\end{eqnarray}%
where $|0\rangle $ is vacuum of bath modes, $\alpha _{k},u_{k_{1}k_{2}},...$
are the coefficients, and even parity is considered. However, it is
impossible to get reasonable results by performing direct  diagonalization in this Fock space
 because of the huge Hilbert-space.
Alternatively, the wavefunction (\ref{Fock}) can be expressed in terms of
another set of complete orthogonal basis, $D\left( \alpha _{k}\right) {\prod
}a_{k_{i}}^{\dagger }|0\rangle $, where $D\left( \alpha _{k}\right) =\exp %
\left[ \sum_{k}\alpha _{k}\left( a_{k}^{\dagger }-a_{k}\right) \right] $ is an
unitary operator , as
\begin{eqnarray}
&&|\Psi \rangle =D\left( \alpha _{k}\right) \left(
1+\sum_{k_{1},k_{2}}b_{k_{1}k_{2}}a_{k_{1}}^{\dagger }a_{k_{2}}^{\dagger
}+...\right) |0\rangle |\uparrow \rangle  \notag \\
&&+D\left( -\alpha _{k}\right) \left(
1+\sum_{k_{1},k_{2}}b_{k_{1}k_{2}}a_{k_{1}}^{\dagger }a_{k_{2}}^{\dagger
}+...\right) |0\rangle |\downarrow \rangle ,  \label{DFS}
\end{eqnarray}%
where the linear term $a_{k}^{\dagger }|0\rangle $ can be  omitted because
the  expansion of the bosonic state  completely
reproduces the first two terms of  Eq. (\ref{Fock}). Note above that the
phonon state in each level is generated by operating on the Fock state with
a unitary displacement operator, called  displaced Fock states
. Only the first term $D\left( \pm \alpha _{k}\right) |0\rangle $ can
reach the whole Hilbert-space, so no truncation is made in this sense. If
the expansion is taken to the infinite order, an exact solution would be
obtained.  However, it is impossible to really perform an infinite order
expansion. Fortunately, it will be shown later that only a few terms in the
expansion would give  very accurate results. In some crucial issues,
converging results can  actually be  achieved.

As a zero-order approximation, we only consider the first term in Eq.
(\ref{DFS}). Projecting the Schr\"{o}dinger equation onto the orthogonal
states $\left\langle 0\right| \ D^{\dagger }\left( \alpha _k\right) \;$and$\
\left\langle 0\right| a_kD^{\dagger }\left( \alpha _k\right) $ gives
\begin{equation}
E=\sum_k\omega _k\alpha _k^2+\sum_kg_k\alpha _k-\frac \Delta 2\exp \left[
-2\sum_k\alpha _k^2\right],  \label{Eq_ohmic_01}
\end{equation}
\begin{equation}
\alpha _k=\frac{-\frac 12g_k}{\omega _k+\Delta \exp \left( -2\sum_k\alpha
_k^2\right) }. \label{Eq_dis}
\end{equation}
Solving $\alpha _k$ self-consistently we can obtain the GS energy and wavefunction immediately. It is interesting to note that the above two equations are  no other than those in the previous SH ansatz  ~\cite{Silbey}.  In other words,
we arrive at the previous well-known analytical results only by the zero-order approximation.

What is more,  anchored to the SH one, the present ansatz for the wavefunction  ({\ref{DFS}}) can be improved by adding more displaced Fock states straightforwardly.
Since the linear term is absent, we can perform the second-order correction by
keeping first two terms. Note that the correlations
between two bosons are fully built in. Analogously, projecting the Schr\"{o}dinger equation onto $%
\left\langle 0\right\vert \ D^{\dagger }\left( \alpha _{k}\right)
,\;\left\langle 0\right\vert a_{k}D^{\dagger }\left( \alpha _{k}\right) $ ,
and $\ \left\langle 0\right\vert a_{k_{1}}a_{k_{2}}D^{\dagger }\left( \alpha
_{k}\right) $ yields the following three equations for unknown $E,\alpha
_{k} $, and$\ b_{k_{1},k_{2}}$,
\begin{equation}
E=\sum_{k}\left( \omega _{k}\alpha _{k}^{2}+g_{k}\alpha _{k}\right) -\frac{1%
}{2}\Delta \eta \left( 1+4\sum_{k}B_{k}\alpha _{k}\right) ,\
\label{Eq_2_en}
\end{equation}%
\begin{equation}
\alpha _{k}=-\frac{\frac{g_{k}}{2}+2\sum_{k^{\prime }}b_{k,k^{\prime }}\left[
\left( \omega _{k^{\prime }}-\Delta \eta \right) \alpha _{k^{\prime }}+\frac{%
g_{k^{\prime }}}{2}\right] }{\omega _{k}+\Delta \eta \left(
1+4\sum_{k}B_{k}\alpha _{k}\right) },  \label{Eq_21}
\end{equation}%
\begin{equation}
b_{k_{1},k_{2}}=-\frac{B_{k_{1}}\alpha _{k_{2}}+B_{k_{2}}\alpha
_{k_{1}}-\alpha _{k_{1}}\alpha _{k_{2}}\left( 1+4\sum_{k}B_{k}\alpha
_{k}\right) }{2\sum_{k}B_{k}\alpha _{k}+\left( \omega _{k_{1}}+\omega
_{k_{2}}\right) /\left( \Delta \eta \right) },  \label{Eq_22}
\end{equation}%
where
\begin{equation}
B_{k}=\sum_{k^{\prime }}b_{k,k^{\prime }}\alpha _{k^{\prime }},\eta =\exp
\left[ -2\sum_{k}\alpha _{k}^{2}\right].
\end{equation}%
$\alpha _{k}$ and $b_{k_{1},k_{2}}$ can be obtained by solving the two coupled  Eqs. (\ref{Eq_21}) and (\ref{Eq_22}) self-consistently, and therefore all observables can be calculated.

Note  that it  is a non-perturbative many-body approach. In both the zeroth- and
second-order approximations, the contributions from infinite Feynman diagrams are
essentially contained, as expected from the iteration spirit in the
self-consistent calculations.

Proceeding as the scheme outlined above, we can straightforwardly perform
the further expansion in the orthogonal displaced Fock basis $D\left( \alpha
_{k}\right) {\prod }a_{k_{i}}^{\dagger }|0\rangle $ systematically step by
step, and get the solution within any desired accuracy in principle. The
challenges remain on the pathway to the high dimensional integral in the
high order study, due to both the analytical derivations and exponentially
increasing computational difficulties. In this paper, we are able to   perform
the corrections up to the fourth-order practically.

\begin{figure}[tbp]
\includegraphics[scale=0.4]{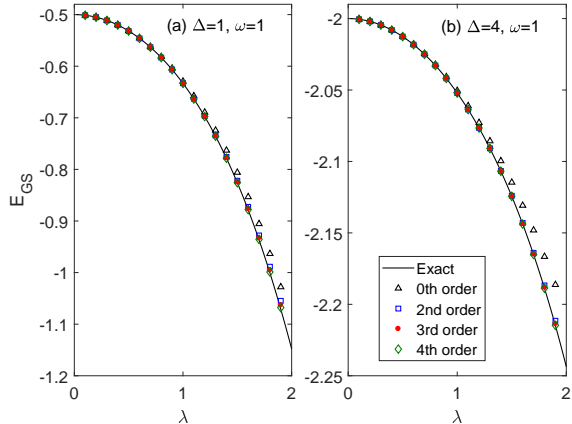}
\caption{ (Color online) Comparisons of GS energy in the zeroth order,
second-order,  third-order, and fourth-order approximations with the numerically exact
ones (solid lines).}
\label{rab_GS}
\end{figure}

To demonstrate  the effectiveness of this approach, we test it in the
single-mode spin-boson model, i.e. the quantum Rabi model where the exact GS
is known \cite{Chen2012}. We present the GS energy within four different  order approximations and
compare with the exact ones in Fig. \ref{rab_GS}. It is interesting to note
that the GS energy in high order
ansatz converges rapidly to the exact ones.  It follows that a finite order approximation in the present ansatz can yield sufficiently accurate results.

\section{Accurate results in  the spin-boson model}

Then we apply the new method  to the spin-boson model.
We stress here that in the present approach we do not have to discretize the
bosonic energy band like in many previous studies at the very beginning. All
$k-$summations involved in the derived equations can be transformed into
continuous integrals like $\int_{0}^{\omega _{c}}d\omega J(\omega )I(\omega )
$. It can be numerically calculated within a Gaussian-logarithmical
integration with very high accuracy, as described in Appendix A in
detail. Without loss of generality, we set $\Delta =0.1,\omega _{c}=1$ in
the calculation if not specified.

\begin{figure}[tbp]
\includegraphics[scale=0.6]{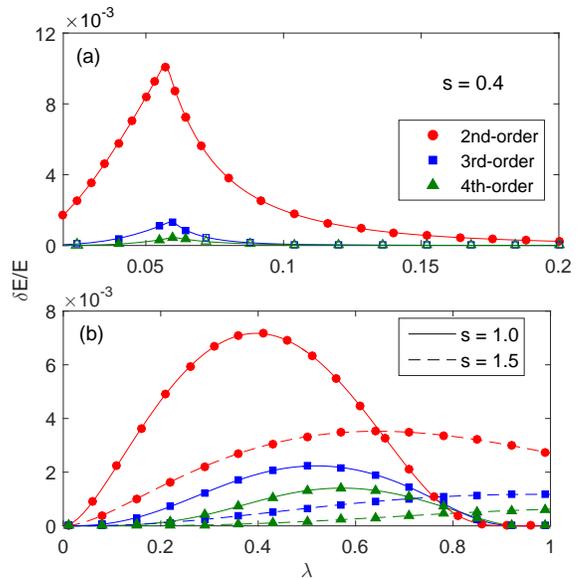}
\caption{ (Color online) The relative difference of the ground state energy
in successive ansatz $\protect\delta E/E$ as a function of $\protect\lambda$
in  the second-, third-, and fourth-order approximations for both exponents (a) $s=0.4$ and
(b) $s=1$ and $1.5$.}
\label{SB_GS}
\end{figure}

For the sub-Ohmic baths, due to the QPT from the delocalized phase to the localized
one, we should relax wavefunction (\ref%
{DFS}) to the asymmetrical one straightforwardly. The zeroth-order
approximation is obviously the same as the asymmetric SH ansatz ~\cite{Chin}. The derivation up to the second-order is given in
Appendix B. For the Ohmic and super-Ohmic spin-boson model, we still
employ the symmetrical ansatz of the wavefunction for $\lambda <1$  where no phase transitions occur.

In Fig. ~\ref{SB_GS}, we present the results for the relative difference of
the GS energy $\delta E/E=(E_{i\text{th}}-E_{(i-1)\text{th }})/E_{(i-1)\text{th}}$ in the successive order
for three typical bath exponents $s=0.4$, $1$, and $1.5$. It is found that the successive corrections
decrease monotonously and tend to convergence very quickly in all cases. The relative difference for the GS energy in  the fourth and third-order approximations is only around $10^{-4}$.

\begin{figure}[tbp]
\includegraphics[scale=0.6]{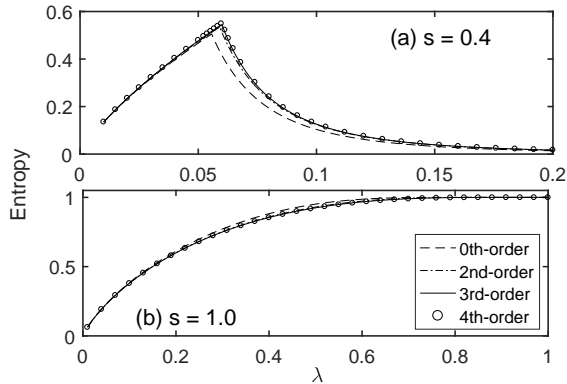}
\caption{ (Color online) The entropy of the ground-state for both $s=0.4$ and  $s = 1.0$ with the zeroth, second,third and fourth order approximations. }
\label{SB_entropy}
\end{figure}

We also calculate the entanglement entropy between the qubit and the bath.
In the spin-boson model, entanglement entropy can be obtained as~\cite{Kopp}:
\begin{equation*}
S=-p_{+}\log p_{+}-p_{-}\log p_{-},
\end{equation*}%
where $p_{\pm }=\frac{1}{2}\left( 1\pm \sqrt{\langle \sigma _{x}\rangle ^{2}+\langle
\sigma _{z}\rangle ^{2}}\right).$ The entanglement entropy as a function
of the coupling strength is given in Fig. ~\ref{SB_entropy} for $s=0.4$ and $%
s=1\;$. With successively higher order study, the
entanglement entropy converges very quickly. The entanglement entropy exhibits a cusp for $s=0.4$, which can be used to locate the critical points of the quantum phase transition.
For the Ohmic bath, the converging entropy increases monotonically with $\lambda$, and  saturates at $S\approx 1$ for $\lambda>1/2$, in excellent agreement with previous results using
the Bethe ansatz solutions~\cite{Kopp}.

\begin{figure}[tbp]
\includegraphics[width=8cm]{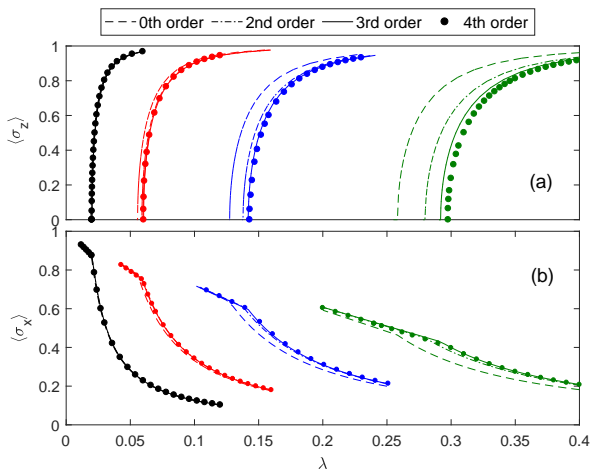}
\caption{(Color online) The magnetization $\langle \sigma_z\rangle$ (a) and  $\langle \sigma_x\rangle$  (b) as a function of the coupling
strength up to the fourth-order asymmetric SH ansatz for $s=0.2,0.4,0.6$, and $0.8$ (from left to right).}
\label{magnetization}
\end{figure}
\textsl{Quantum phase transition for sub-Ohmic baths.-}. For the sub-Ohmic
bath, the rich physics of the quantum dissipation is second-order QPT from
delocalization to localization as a consequence of the competition between
the amplitude of tunneling of the spin and the effect of the dissipative
bath. The magnetization $M=\langle \sigma_z\rangle$ can be used as an order parameter in the QPT of
this model. We calculate the
magnetization $M$ and evaluate the critical coupling strength $\lambda _{c}$
up to the fourth-order ansatz.

The results of  magnetization for $s=0.2,0.4,0.6$, and $0.8$
are exhibited in Fig. ~\ref{magnetization}(a). In each order approximation,
there obviously exists a critical point which separates the localized one ($%
M\neq 0$) from the delocalized phase ($M=0$). Interestingly, both $M$ and $\lambda _{c}$ display a fast converging behavior. Especially for  bath exponents $s=0.2$ and $%
s=0.4$, the converging magnetization has been already achieved. We believe that  the almost  exact GS results in the bath
exponent regime $0<s<1/2$  can be obtained in the present approach up to the fourth-order ansatz.

However, for $s > 1/2$, the convergence is not arrived at until
the fourth-order calculations. It is expected that the more accurate results
could be obtained in the further corrections, which is however
prohibitively expensive in the high dimensional continuous integral. As
found recently by Blunden-Codd \textsl{et al.},  \cite{Blunden},  the quantum
criticality associated to an
interacting fixed point \cite{Vojta1} in this nontrivial regime  can be only touched by the very
accurate wavefunction where  with at least a hundred of single-mode coherent states are  required, which is beyond the scope of the present study.

The renormalized tunneling matrix element  $\langle \sigma_x\rangle$ in the GS for different value of $s$ is also presented in  Fig. ~\ref{magnetization}(b). A fast converging behavior is also found. Note that,  after the critical point, it decreases more rapidly, but does not vanish.

Both the magnetization and entanglement entropy can give
the same $\lambda _{c}$. For convenience, we list critical coupling
strengths by the present different order approximations in Table I. The
corresponding QMC results, which are drawn from  Ref. \cite{QMC}, are also collected for
comparison. $\lambda _{c}$ generally  increases  with the increasing order. Convergency is achieved obviously for $s<1/2$.  Interestingly, $\lambda _{c}=0.0600$ for $\ s=0.4$ in our
fourth-order study is almost identical to  the previous QMC one $0.0601$ ~\cite{QMC}.

\begin{widetext}
\begin{center}
\begin{table}[h]
\caption{The critical coupling strengths within different  order approximations are collected. The sixth column presents those by quantum Monte Carlo simulations~\cite{QMC}.}
\begin{tabular}{p{2.0cm}p{2.0cm}p{2.0cm}p{2.0cm}p{2.0cm}p{2.0cm}}
\hline\hline
$s$ & $0th$ & $2nd$ & $3rd$ & $4th$  & $QMC^{Ref. [21]}$  \\ \hline
$0.2$ & 0.0195 & 0.02005 & 0.02013 & 0.02014 &  0.0175  \\
$0.4$ & 0.0557 & 0.0590 &0.0599 & 0.0600 & 0.0601  \\
$0.6$ & 0.127& 0.138 & 0.142 & 0.143 & 0.155  \\
$0.8$ & 0.258 &  0.280 & 0.292 & 0.297 & 0.359  \\ \hline\hline
\end{tabular}
\end{table}
\end{center}
\end{widetext}
Discretization of the energy spectrum of the continuum bath should be performed at the very
beginning in many  advanced numerical approaches~\cite{Bulla,vojta,phi,Tong,ED, Zhang,DMRG,VMPS,Frenzel,Kirchner}, except in the QMC where the bath is integrated out. The artificial
discretization definitely yields quantitative calculation errors. Only QMC simulations may in principle provide reliable and unique  critical points with high
accuracy.   One may note that our
result for $\lambda _{c}$ for $s=0.2$ is close to but a little bit larger
than the QMC one. It unlikely converges to the QMC value due to the trend of
convergence. The statistical error in QMC simulations may account for this
slight difference.

\begin{figure}[tbp]
\includegraphics[width=8cm]{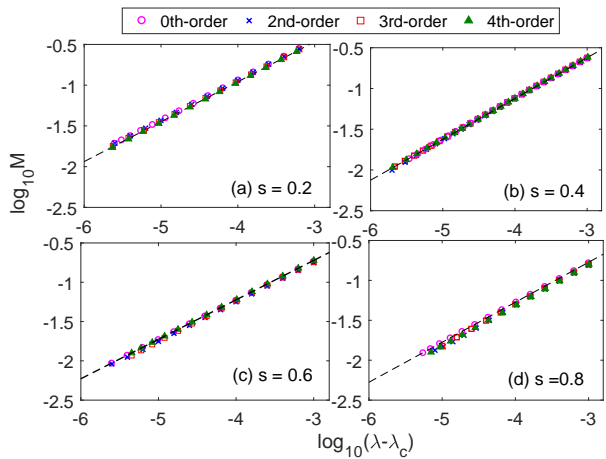}
\caption{ (Color online) The log-log plot of the magnetization $M$ as a
function of $\protect\lambda -\protect\lambda _{c}$ for $s=0.2,0.4,0.6$, and $0.8$. The power
law curves with $\protect\beta =0.5$ are denoted by the dashed line.}
\label{critical_curves12}
\end{figure}

In the second-order QPT, the order parameter should  display a power law behavior near the critical point $M\propto
\left( \lambda -\lambda _{c}\right) ^{\beta }$.  We  present the magnetization within the second-, third-order and fourth-order ansatz as a function
of $\lambda -\lambda _{c}$ in a log-log plot for $s=0.2,0.4,0.6$, and $0.8$ in Fig. \ref%
{critical_curves12}. It is observed that the curves almost coincide in the whole
coupling regime. A very nice power-law behavior  over  two decades
with an exponent $\beta =0.5\pm 0.01$ is demonstrated for all cases. Therefore even the fourth-order study does not modify the exponent $\beta $
either, indicating a robust mean-field nature in the regime $s<1/2$ of this model.

\textsl{Biased spin-boson model.-} Extension to the biased spin-boson model
can be performed straightforwardly. In this case, a static bias term $\frac \epsilon 2\sigma _z$ is added to Hamiltonian (\ref{hamiltonian}).
Obviously, the
asymmetrical wavefunction should be entailed in the asymmetrical spin-boson
model with any baths. Nazir et al. found that the GS magnetization ($M$) within the SH ansatz jumps to $-1$, i.e. the value for the fully localized state, for some value of $\lambda$ ~\cite{Nazir}, whereas many advanced approaches have explicitly led to the smooth crossover behavior.

\begin{figure}[tbp]
\includegraphics[width=8cm]{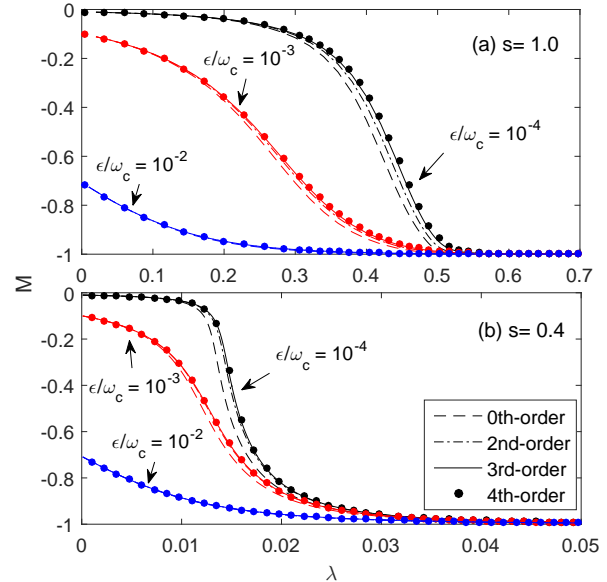}
\caption{ (Color online) The log-log plot of the magnetization $M$ as a
function of $\protect\lambda -\protect\lambda _{c}$ for (a) $s=1.0$, and (b)
$s=0.4$. The power law curves with $\protect\beta =0.5$ are denoted by the
dashed line.}
\label{bias}
\end{figure}

We set the same parameters as in Ref. \cite{Nazir}, $\Delta /\omega _{c}=0.01$,
and also calculate $M$ as a function of $\lambda$ within the present ansatz. As shown in Fig. \ref{bias} that, for the Ohmic baths, the unphysical discontinuous "jump" for
small bias $\epsilon $ does not appear in any order
approximations. We believe that the previous surprising jump is caused by the symmetrical SH variational wavefunction used in Ref. \cite{Nazir}. Moreover, fast convergence of the magnetization is shown in  Fig. \ref{bias}  for both $s=1$ and $0.4$.  Especially, for $s=0.4$,
a converging result is arrived even  in the third-order ansatz. It is  demonstrated that the present approach is also quite efficient in the biased case.

\section{Conclusion}

Previous well-known SH polaron ansatz in the spin-boson model is
improved successively in this paper. The GS wavefunction can be written in
the expansion of orthogonal displaced Fock states. Projecting the Schr\"{o}%
dinger equation onto orthogonal displaced Fock states with increasing order gives a set of coupled equations. Solving the equations
self-consistently allows for the GS observables. The zeroth-order one is just the same as the  SH ansatz.
More importantly, it can be  easily extended to the high order, which is actually the expansion  anchored around SH ansatz. More accurate results can  then  be achieved
with increasing order where correlations among more bosons are included. In
principle, the exact results could be archived in the infinite-order study,
which is, however, very challenging at the moment. Fortunately, very accurate
results can be obtained in a few higher order calculations that we can
performed practically. Especially in the sub-Ohmic spin-boson model, the
fourth-order ansatz can yield the converging results for $0<s<1/2$.  The present approach is also suited to the biased spin-Boson model.
 It is expected that this approach paves the way to the true exact solution
analytically. Generalizations to quantum dynamics at both zero and finite temperatures are  progressing.

\appendix

\section{Gaussian-logarithmical integration for the continuous integral}

Here we demonstrate how to transform the $k-$summation appearing in Eqs. (5-10) into the continuous integral and then illustrate
an effective numerical calculation with high accuracy.

In the zero-order
approximation, also the well known SH ansatz, we can set
\begin{equation*}
\alpha _k=\alpha _k^{\prime }g_k,
\end{equation*}
Eq. (\ref{Eq_dis}) then becomes
\begin{equation*}
\alpha _k^{\prime }=-\frac{1/2}{\omega _k+\Delta \exp \left( -2\sum_k\alpha
_k^{\prime 2}g_k^2\right) },
\end{equation*}
so $\alpha _k^{\prime }$ is only related to $g_k$ implicitly.

In the second-order approximation, we can set
\begin{eqnarray*}
\alpha _k &=&\alpha _k^{\prime }g_k, \\
b_{k_1,k_2} &=&b_{k_1,k_2}^{\prime }g_{_{k_1}}g_{_{k_2}}.
\end{eqnarray*}
Inserting to Eqs. (\ref{Eq_21}) and (\ref{Eq_22}) gives
\begin{equation*}
\alpha _k^{\prime }=\frac{-\frac 12+2\sum_{k^{\prime }}g_{k^{\prime
}}^2b_{k,k^{\prime }}^{\prime }\left[ \left( \omega _{k^{\prime }}-\Delta
\eta \right) \alpha _{k^{\prime }}^{\prime }+1/2\right] }{\omega _k+\Delta
\eta \left( 1+4\zeta \right) },
\end{equation*}

\begin{widetext}
\begin{equation*}
b_{k_1,k_2}^{\prime }=\frac{\alpha _{k_1}^{\prime }\alpha _{k_2}^{\prime
}\left( 1+4\zeta \right) -\sum_{k^{\prime }}g_{k^{\prime }}^2\alpha
_{k^{\prime }}^{\prime }\left( b_{k_1,k^{\prime }}^{\prime }\alpha
_{k_2}^{\prime }+b_{k_2,k^{\prime }}^{\prime }\alpha _{k_1}^{\prime }\right)
}{2\zeta +\left( \omega _{k_1}+\omega _{k_2}\right) /\left( \Delta \eta
\right) },
\end{equation*}
\end{widetext}
where
\begin{equation*}
\zeta =\sum_kg_k^2\sum_{k^{\prime }}g_{k^{\prime }}^2b_{k,k^{\prime
}}^{\prime }\alpha _{k^{\prime }}^{\prime }\alpha _k^{\prime }.
\end{equation*}
Given $g_k$, both $\alpha _k^{\prime }$ and $b_{k_1,k_2}^{\prime }\;$can be
obtained self-consistently. Note that each $k$-summation takes the form of $%
\sum_kg_k^2I(k)$ where $I(k)$ does not depend on $g_k^2\;$explicitly, and so
both $\alpha _k^{\prime }$ and $b_{k_1,k_2}^{\prime }\;$are functionals of $%
g_k$. $\;$Without loss of generality, $k$ is corresponding to $\omega $ one
by one, the $k$-summation can be transformed to the $\omega \;$integral as
\begin{equation*}
\sum_kg_k^2I(k)\rightarrow \int_0^{\omega _c}d\omega \frac{J(\omega )}\pi
I(\omega ),
\end{equation*}
so we have
\begin{equation}
\alpha ^{\prime }(\omega )=\frac{-\frac 12+\xi (\omega )-2\Delta \eta \chi
(\omega )}{\omega +\Delta \eta \left( 1+4\zeta \right) },
\label{displacement}
\end{equation}

\begin{equation}
b^{\prime }\left( \omega _1,\omega _2\right) =\frac{\alpha ^{\prime }(\omega
_1)\alpha ^{\prime }(\omega _2)\left( 1+4\zeta \right) -\kappa (\omega
_1,\omega _2)}{2\zeta +\left( \omega _{_1}+\omega _{_2}\right) /(\Delta \eta
)},  \label{sec_coeff}
\end{equation}
where
\begin{eqnarray*}
\xi (\omega ) &=&\int_0^{\omega _c}d\omega ^{\prime }\frac{J(\omega ^{\prime
})}\pi \left[ 2\omega ^{\prime }\alpha ^{\prime }(\omega ^{\prime })+1\right]
b^{\prime }\left( \omega ,\omega ^{\prime }\right) , \\
\chi (\omega ) &=&\int_0^{\omega _c}d\omega ^{\prime }\frac{J(\omega
^{\prime })}\pi \alpha ^{\prime }(\omega ^{\prime })b^{\prime }\left( \omega
,\omega ^{\prime }\right) , \\
\kappa (\omega _1,\omega _2) &=&\chi (\omega _1)\alpha ^{\prime }(\omega
_2)+\chi (\omega _2)\alpha ^{\prime }(\omega _1),
\end{eqnarray*}
are some functions of $\omega ,$ and
\begin{eqnarray*}
\zeta &=&\int_0^{\omega _c}d\omega \frac{J(\omega )}\pi \int_0^{\omega
_c}d\omega ^{\prime }\frac{J(\omega ^{\prime })}\pi \alpha ^{\prime }(\omega
)\alpha ^{\prime }(\omega ^{\prime })b^{\prime }\left( \omega ,\omega
^{\prime }\right) , \\
\eta &=&\exp \left[ -2\int_0^{\omega _c}d\omega ^{\prime }\frac{J(\omega
^{\prime })}\pi \alpha ^{\prime 2}(\omega ^{\prime })\right] ,
\end{eqnarray*}
are constants.

The GS energy (\ref{Eq_2_en}) can be expressed  as
\begin{equation}
E=\int_0^{\omega _c}d\omega \frac{J(\omega )}\pi \alpha ^{\prime }(\omega )%
\left[ \omega \alpha ^{\prime }(\omega )+1\right] -\frac 12\Delta \eta
\left( 1+4\zeta \right) \ .
\end{equation}

The self-consistent solutions of two coupled equations Eqs. (\ref%
{displacement}) and (\ref{sec_coeff}) are in no way obtained analytically,
numerical calculations should be performed. Note that the low frequency modes
play the dominant role in the QPT of the sub-ohmic spin-boson model. There
is an infrared divergence of the integrand like $\int_{0}^{\omega
_{c}}\omega ^{s-2}d\omega $ in the limit of $\omega \rightarrow 0$ for the
sub-Ohmic and Ohmic baths, which is called as the infrared catastrophe. Thanks to the
Gaussian quadrature rules, where the zero frequency is not touched. We can
discretize the whole frequency interval with Gaussian grids, the integral
can be numerically exactly achieved with a large number of Gaussian grids.
It is very time consuming to calculate the integral in this way, especially
for high dimensional integral involved in the high order approximation. According to
the structure of the integrand, it is not economical to deal with the high
and low frequency regime on the equal footing.

 To increase the efficiency,
we combine the logarithmic discretization with Gaussian quadrature rule~\cite
{Richard}. First, we divide the $\omega $ interval $[0,1]$ into $L+1$
sub-intervals as $[\Lambda ^{-(l+1)},\Lambda ^{-l}]\;(l=0,1,2,L-1)\;$and$%
\;[0,\Lambda ^{-L}]$ , then we apply the Gaussian quadrature rule to each
logarithmical sub-interval. So the continuous integral is calculated by the
following summation
\begin{equation}
\int_{0}^{1}J(\omega )I(\omega )d\omega
=\sum_{l=0}^{L}\sum_{n=1}^{N}W_{l,n}J(\omega _{l,n})I(\omega _{l,n}),
\label{combination}
\end{equation}%
where $N$ is the number of gaussian points inserted in each sub-interval, $%
W_{l,n}$ is corresponding Gaussian weight. After careful examination,\ we
find that the Gaussian-logarithmical integration for  integrand in this problem
converges if set $L=6,N=9,$ and $\Lambda =9$, which are used in this work.

\section{Formalism for sub-Ohmic baths}

In the sub-Ohmic spin-boson model, the GS wavefunction can be
generally expressed in the orthogonal displaced Fock basis $D\left( \alpha
_k\right) {\prod }a_{k_i}^{\dagger }|0\rangle $ as
\begin{widetext}
\begin{eqnarray}
&&|\Psi \rangle =D\left( \alpha _k\right) \left(
1+\sum_{k_1,k_2}b_1(k_1,k_2)a_{k_1}^{\dagger }a_{k_2}^{\dagger
}+\sum_{k_1,k_2,k_3}c_1(k_1,k_2,k_3)a_{k_1}^{\dagger }a_{k_2}^{\dagger
}a_{k_3}^{\dagger }+...\right) |0\rangle |\uparrow \rangle  \notag \\
&&+D(\beta _k)\left( r+\sum_{k_1,k_2}b_2(k_1,k_2)a_{k_1}^{\dagger
}a_{k_2}^{\dagger }+\sum_{k_1,k_2,k_3}c_2(k_1,k_2,k_3)a_{k_1}^{\dagger
}a_{k_2}^{\dagger }a_{k_3}^{\dagger }+...\right) |0\rangle |\downarrow
\rangle,  \label{sub-Ohmic}
\end{eqnarray}
\end{widetext}
where $r, \alpha _k, \beta _k$, and $b_i, c_i (i=1,2)$ are asymmetrical
coefficients to be determined.

In the zero-order approximation, we only select the first term in Eq. (\ref%
{sub-Ohmic}). Projecting the Schr\"{o}dinger equation in the upper level
onto the orthogonal basis $\left\langle 0\right| \ D^{\dagger }\left( \alpha
_k\right)$ and $\ \left\langle 0\right| a_kD^{\dagger }\left( \alpha
_k\right) $ and in the lower level onto $\left\langle 0\right| \ D^{\dagger
}\left( \beta _k\right) \;$and$\ \left\langle 0\right| a_kD^{\dagger }\left(
\beta _k\right) $ results in
\begin{eqnarray}
\sum_k\left( \omega _k\alpha _k^2+g_k\alpha _k\right) -\frac \Delta 2r\
\Gamma &=&E,  \label{E_upper} \\
\omega _k\alpha _k+\frac 12g_k+\frac \Delta 2r\Gamma D_k &=&0,
\label{dis_upper}
\end{eqnarray}
and
\begin{eqnarray}
\sum_k\left( \omega _k\beta _k^2-g_k\beta _k\right) -\frac \Delta {2r}\
\Gamma &=&E,  \label{E_down} \\
\omega _k\beta _k-\frac 12g_k-\frac \Delta {2r}\Gamma D_k &=&0,
\label{dis_down}
\end{eqnarray}
where
\begin{eqnarray*}
\Gamma &=&\exp \left[ -\frac 12\sum_kD_k^2\right] , \\
D_k &=&\alpha _k-\beta _k,
\end{eqnarray*}
which are the same as those obtained variationally within the generalized SH
ansatz \cite{Chin}.

In the second-order approximation, the first two terms in each level of Eq. (\ref{sub-Ohmic}) are
kept. Proceeding as procedures outlines above, projecting the Schr\"{o}%
dinger equation in the upper level onto the orthogonal states $\left\langle
0\right| \ D^{\dagger }\left( \alpha _k\right) ,\ \left\langle 0\right|
a_kD^{\dagger }\left( \alpha _k\right) $, and $\ \left\langle 0\right|
a_{k_1}a_{k_2}D^{\dagger }\left( \alpha _k\right) $ and in the lower level
onto $\left\langle 0\right| \ D^{\dagger }\left( \beta _k\right)
,\left\langle 0\right| a_kD^{\dagger }\left( \beta _k\right)$, and $\
\left\langle 0\right| a_{k_1}a_{k_2}D^{\dagger }\left( \beta _k\right) $
yield the following six equations

\begin{widetext}
\begin{eqnarray}
\sum_k\left[ \omega _k\alpha _k^2+g_k\alpha _k\right] -\frac \Delta 2\Gamma %
\left[ r+\sum_kB_kD_k\right] \ &=&E,  \label{sub_1up} \\
r\sum_k\left[ \omega _k\beta _k^2-g_k\beta _k\right] -\frac \Delta 2\Gamma %
\left[ 1+\sum_kA_kD_k\right] \ &=&rE,  \label{sub_1down}
\end{eqnarray}

\begin{equation}
\left[ \omega _k\alpha _k+\frac{g_k}2\right] +\sum_{k^{\prime }}2b_1\left(
k,k^{\prime }\right) \left[ \omega _{k^{\prime }}\alpha _{k^{\prime }}+\frac{%
g_{k^{\prime }}}2\right] -\Delta \Gamma B_k+\frac \Delta 2\Gamma D_k\left[
r+\sum_kB_kD_k\right] =0,  \label{sub_2up}
\end{equation}
\begin{equation}
r\left[ \omega _k\beta _k-\frac{g_k}2\right] +\sum_{k^{\prime }}2b_2\left(
k,k^{\prime }\right) \left[ \omega _{k^{\prime }}\beta _{k^{\prime }}-\frac{%
g_{k^{\prime }}}2\right] +\Delta \Gamma A_k-\frac \Delta 2\Gamma D_k\left[
1+\sum_kA_kD_k\right] =0,  \label{sub_2down}
\end{equation}
\begin{eqnarray}
&&b_1(k_1,k_2)\left( \omega _{k_1}+\omega _{k_2}\right) +\frac \Delta
2\Gamma \left[ r+\sum_kB_kD_k\right] b_1(k_1,k_2)  \notag \\
&&-\frac \Delta 2b_2(k_1,k_2)\Gamma +\frac \Delta 2\Gamma \left[
B_{k_1}D_{k_2}+B_{k_2}D_{k_1}\right] -\frac \Delta 4\Gamma D_{k_1}D_{k_2}%
\left[ r+\sum_kB_kD_k\right] =0,  \label{sub_3up}
\end{eqnarray}
\begin{eqnarray}
&&b_2(k_1,k_2)\left( \omega _{k_1}+\omega _{k_2}\right) +\frac \Delta
{2r}\Gamma \left[ 1+\sum_kA_kD_k\right] b_2(k_1,k_2)  \notag \\
&&-\frac \Delta 2b_1(k_1,k_2)\Gamma +\frac \Delta 2\Gamma \left[
A_{k_1}D_{k_2}+A_{k_2}D_{k_1}\right] -\frac \Delta 4\Gamma D_{k_1}D_{k_2}%
\left[ 1+\sum_kA_kD_k\right] =0,  \label{sub_3down}
\end{eqnarray}
\end{widetext}
where
\begin{eqnarray*}
\ \ A_k &=&\sum_{k^{\prime }}b_1(k^{\prime },k)D_{k^{\prime }}, \\
\ \ B_k &=&\sum_{k^{\prime }}b_2(k^{\prime },k)D_{k^{\prime }}.
\end{eqnarray*}
The self-consistent solutions for the above coupled equations will give all
results in the second-order study. If set $r=1,\alpha _k=-\beta _k$ and $%
b_1(k_1,k_2)=b_2(k_1,k_2)$, Eqs. (\ref{Eq_21}) and (\ref{Eq_22}) for the
symmetric case are recovered completely.

It is straightforward to perform the third-order and fourth-order studies by further adding the
third and the fourth terms in  Eq. (\ref{sub-Ohmic}). The derivations are  lengthy and not
shown here.

\textbf{ACKNOWLEDGEMENTS}
This paper is supported by the National Science Foundation of China (No. 11474256 and No.  11674285), and the National Key Research and Development Program of China (No.   2017YFA0303002).

$^{\ast }$ Corresponding author. Email:qhchen@zju.edu.cn


\end{document}